\documentclass[journal=jacsat,manuscript=article]{achemso}

\usepackage{chemformula} 
\usepackage[T1]{fontenc} 
\usepackage{xcolor}
\usepackage{physics}
\usepackage{amsmath,bm}
\usepackage{amsmath,MnSymbol}
\usepackage{amsmath}
\usepackage{verbatim}
\usepackage{amsfonts} 

\usepackage{graphicx}
\usepackage{dcolumn}
\usepackage{bm}
\author{Abhineet Singh Rajput}
\alsoaffiliation{$^{\dagger}$Department of Mechanical Engineering, Indian Institute of Science, Bangalore, India}
\author{Sarath Chandra Varma}
\alsoaffiliation{$^{\dagger}$Department of Mechanical Engineering, Indian Institute of Science, Bangalore, India}
\author{Aloke Kumar}
\alsoaffiliation{$^{\dagger}$Department of Mechanical Engineering, Indian Institute of Science, Bangalore, India}
\email{*alokekumar@iisc.ac.in}

\title[An \textsf{achemso} demo]
  {Newtonian coalescence in colloidal and non-colloidal suspensions}

\abbreviations{IR,NMR,UV}

\begin{document}







\begin{abstract}

Coalescence event in pendant and sessile drop is distinguished by the formation and evolution of the liquid bridge created upon singular contact. The bridge radius, $R$, is known to evolve as $R\sim t^b$, with power-law exponent, $b$, signifying the dominant governing forces. In this work, we experimentally explore the phenomenon in  sub-classes of complex fluids namely, colloidal and non-colloidal suspensions that have particle hydrodynamic interactions as origin of viscoelasticity. Our observations suggest that such fluids have flow dependent thinning response with finite elasticity in shear flows but negligible in extensional flows. Based on these, the study extends the Newtonian universality of $b=0.5$ to these thinning fluids. Further we fortify these observations through a theoretical model developed by employing Ostwald-de Waele constitutive law. Finally, we utilize this theoretical model to inspect the existence of arrested coalescence in generalized Newtonian fluids.      

\end{abstract}

\section{Introduction}

Droplet coalescence is a thermodynamic equilibration process driven by surface energy minimization\cite{76}. The physics of this phenomenon is characterized by the temporal evolution of a liquid bridge formed upon the proximate approach of two droplets\cite{16}. This phenomenon is ubiquitous, manifesting in processes linked to life like those in raindrop formation\cite{1,3}, growth of tumor cells\cite{tumor}, and industrial processes like those in combustion\cite{14}, spray paintings\cite{5}, and coatings\cite{6}. Despite these universal occurrences, studies on the coalescence of complex fluid droplets remain scarce in the literature. Unlike Newtonian fluids, complex fluids have signature micro-structures that can result in a wide range of responses depending on external perturbations making a unified model elusive\cite{bird}. Such diversity in micro-structures and flow behaviors have resulted in classifying these in sub-classes ranging from polymers to suspensions. But there has been a recent surge in studies investigating coalescence dynamics in macromolecular-based micro-structure fluids, i.e., polymeric fluids\cite{our,varma2021coalescence,varma2022rheocoalescence,doi:10.1063/5.0112846,rajput2022sub}. However, coalescence in other complex fluid sub-classes like those in suspensions where micro-structures are comprised of particles having hard-sphere interactions remains unknown.  

Depending on the geometry of the problem, literature classifies coalescence primarily in three distinct configurations: pendant-sessile\cite{our,varma2022rheocoalescence,rajput2022sub}, sessile-sessile\cite{varma2021coalescence,doi:10.1063/5.0112846}, and pendant-pendant\cite{18,19,17}. For pendant-sessile and pendant-pendant geometries, the merging kinematics is identified by the liquid bridge of semi-width, $H$, and radius, $R$. Previous studies on Newtonian droplets\cite{18,19,21,67} have demonstrated the bridge radius $R$ to evolve in time, $t$, as $R\sim t^b$ through an interplay of the capillary, viscous and inertial forces. The power-law exponent, $b$, assumes regime dependent universal values of $b=1$ and $b=0.5$ in viscous\cite{18,19} and inertial\cite{21} dominant regimes respectively. However, recent studies on a particular subclass of complex fluids, i.e., gels\cite{CHEN2022283} and polymers\cite{our,varma2022rheocoalescence,rajput2022sub}, have reported a deviation from Newtonian behavior due to the presence of an additional elastic force during coalescence. Our recent studies on macromolecular fluids\cite{our,varma2022rheocoalescence,rajput2022sub} have demonstrated that the power-law exponent $b$ is not always universal; instead, it has a regime-dependent behavior that is classified based on concentration ratio, $c/c*$. The power-law exponent, $b$, assumes a universal value of $b=0.37$ in inertio/viscoelastic regimes while it monotonically decreases in the elasticity-dominated regime. However, the results proposed by these studies cannot be generalized to other sub-classes of complex fluids owing to differences in micro-structures that give rise to non-Newtonian behaviors. In particular, for colloidal and non-colloidal suspensions, the dispersed phases can comprise of rigid particles that only contribute to dissipation, unlike macromolecular phases with both storage and dissipative characteristics.

In this study, we probe the coalescence paradigm in colloidal and non-colloidal suspensions through experiments and theoretical modeling. Based on rheological characterisation, we classify these as power-law fluids, $\eta=\kappa(\dot{\gamma})^n$ (where, $\eta$ is viscosity, $\dot{\gamma}$ is shear rate, $\kappa$ the flow consistency index  and $n$ the degree of non-Newtonian behaviour) with exponent $n<1$, $n=1$ and $n>1$ representing thinning, Newtonian and thickening fluids, respectively, and utilize Ostwald-de Waele constitutive law for developing the theoretical framework. Our experimental results demonstrate the sustenance of Newtonian universality of $b=0.5$ in thinning fluids (both shear\cite{tanner2020computation} and elongational\cite{tanner2020computation}), which is the signature of the difference in the microstructure of the two subclasses, colloidal and non-colloidal suspensions and macromolecular fluids. In macromolecular fluids, chain relaxation designated by the relaxation time $\lambda$ is the signature of viscoelasticity. However, for colloidal and non-colloidal suspensions, viscoelasticity originates via hydrodynamic interactions among particles that are induced through external flow fields. Although such fluids also have finite relaxation time $\lambda$ arising from particle self-diffusion\cite{shikata1994viscoelastic} or inertial relaxation\cite{shaw2003particle}, the two relaxation mechanisms are different, resulting in shear/extensional elasticity in the former and shear elasticity alone in the latter. These observations are fortified by our theoretical model that generalizes the Newtonian universality of power-law exponent $b=0.5$ to thinning fluids $n<1$ with zero extensional elasticity. Further, we employ the framework to investigate the existence of arrested coalescence in generalized Newtonian fluids. The study also provides insights into the merging dynamics of thickening fluids $n>1$ where the coalescence appears to be delayed. 

\section{Materials and methods}

Colloidal suspensions of various volume fractions, $\phi$, are prepared by adding sufficient quantity of silica fumed powder in DI Water. Similarly, the non-colloidal suspensions of different volume fractions are prepared by adding silica glass spheres of corresponding quantities in glycerol. Further, all colloidal and non-colloidal suspensions are agitated in a shaker at 130 RPM for atleast 6 hrs followed by a sonication for 10 mins to ensure homogeneity.The details of particle diameter, $D_p$, particle density, $\rho_p$, and the manufacturer for the dispersed phase is mentioned in Table-I.  The volume fraction, $\phi$, is calculated as stated in Eq(1) with $m_p$ and $v_s$ being mass of dispersed phase and volume of solvent respectively. Values of $\phi$ are given in Table-II. Polymer solution of concentration, $c$, of $0.1\%$ g/ml is prepared by adding $0.05$ g of Poly(ethylene oxide) $M_w=5\times10^6$ g/mol in $50$ ml of DI water and stirring the solution at 300 rpm for atleast $24$ hrs.  

\begin{equation}
    \phi=\frac{\frac{m_p}{\rho_p}}{\frac{m_p}{\rho_p}+v_s}
\end{equation}

\begin{table}[hbt!]
\caption{\label{tab:table1} List of dispersed phase with their respective diameters and densities used in present study.}

\begin{tabular}{c|c|c|c|c}
Suspension & Dispersed Phase & $D_p$ $\mu$m & $\rho_p$ g/ml & \text{Manufacturer}\\
\hline
Colloidal&Silica fumed&$ 0.2-0.3 $& 2.2 &\text{Sigma-Aldrich}\\
Non-colloidal&Silica glass spheres&$9-13$& 1.1 & \text{Sigma-Aldrich}\\
\end{tabular}

\end{table}

Preceding the experiments PDMS substrates are prepared by cleaning the glass slides with detergent and sonicating them in acetone and water for 20 mins each. These cleansed slides are then dried in a hot air oven at 90$^\circ$C for at least 30 mins. Parallelly,  PDMS is prepared by mixing it with curing agent (Syl Gard 184 Silicone Elastomer Kit, Dow Corning) in the ratio of 10:1. The mixture is then desiccated for 30 minutes until all visible gas bubbles are removed.  Finally, the slides are coated with PDMS using a spin coater by spinning at 5000 rpm for 60 s followed by curing in a hot air oven at 60$^\circ$C for at least 8 hrs.

To perform capillary breakup extensional rheometry dripping on a substrate (CABER-DOS)\cite{dinic2017pinch} experiments  a glass substrate (Blue Star, India) of dimensions $75\times25\times1. 45$ mm is used. Substrates are cleansed with detergent and then sonicated with acetone and water for 20 minutes each. Subsequently, they are placed in the oven at 95$^\circ$C for 30 minutes.

\begin{table}[h!]
  \caption{Rheological properties of the solutions used in present study. }
  \label{tbl:example}
  \begin{tabular}{ll|ll}
    \hline
    ~~~~~Colloidal suspension  &~~~~~~& ~~~~~Non-colloidal suspension \\
    \hline
    ~~~$\phi$ ~~~~~& $\eta_{\infty}$ (mPas) & ~~~$\phi$ ~~~~~& $\eta_{\infty}$(mPas)\\
    \hline
     ~~~DI Water ~~~~~& 1 & ~~~Glycerol ~~~~~&760 \\
     ~~~0.00045 ~~~~~& 1 & ~~~0.0009 ~~~~~& 788\\
     ~~~0.0009 ~~~~~&  1 & ~~~0.009 ~~~~~& 793\\
     ~~~0.0018 ~~~~~& 1 & ~~~0.017 ~~~~~& 830\\
     ~~~0.0045 ~~~~~& 1 & ~~~0.035 ~~~~~& 887\\
     ~~~0.0090 ~~~~~& 1.6 & ~~~0.068 ~~~~~& 875\\
     ~~~0.018 ~~~~~& 3.4 & ~~~0.083 ~~~~~& 955\\
     ~~~0.038 ~~~~~& 16 & ~~~0.15 ~~~~~& 1359\\
     ~~~0.043 ~~~~~& 34 & ~~~0.27 ~~~~~& 2176\\
     ~~~0.051 ~~~~~& 68 & ~~~0.31 ~~~~~& 2676\\
     ~~~ - ~~~~~& - & ~~~0.35 ~~~~~& 3300\\
     \hline

  \end{tabular}
\end{table}

\section{Rheology}
\begin{figure*}[h!]
\includegraphics[width=\linewidth]{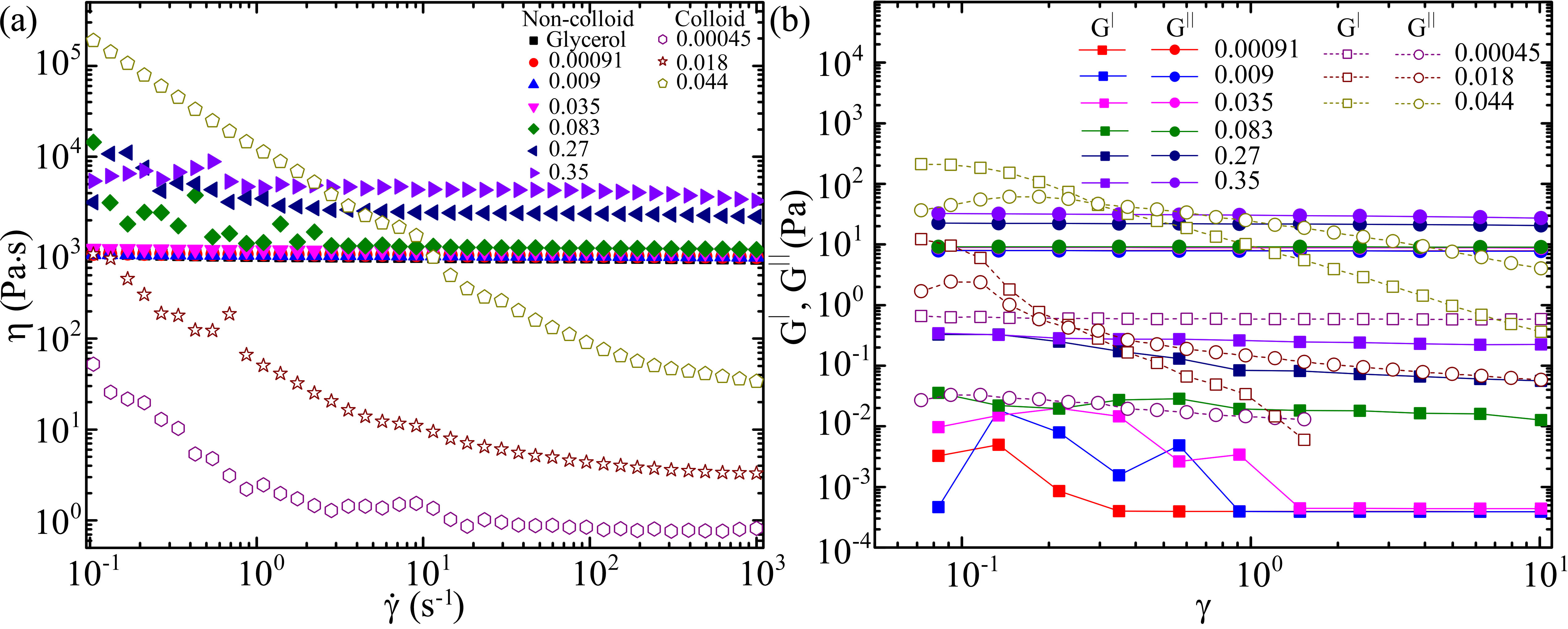}
\centering
\caption{(a) Variation of viscosity with shear rate for various volume fractions of colloidal and non-colloidal suspensions, (b) Amplitude sweep of various volume fractions of colloidal and non-colloidal suspensions to show the degree of viscoelastic behaviour of the solutions.}
\label{fig:Rheometry} 
\end{figure*}

Viscoelastic characterization of colloidal and non-colloidal suspensions are done by performing rheological experiments on cone-plate $40$ mm, 1$^{\circ}$ geometry of Anton Paar$^{\tiny{\text{\textregistered}}}$ MCR 302 rheometer. The dependence of viscosity $\eta$ on shear rates $\dot{\gamma}$ for different volume fractions $\phi$ of colloidal and non-colloidal suspensions are given in Fig.~\ref{fig:Rheometry}a. Colloidal suspensions have shown strong shear-thinning behaviour with $\phi=0.054$ approaching close to sol-gel\cite{kawaguchi2020dispersion} transition while non-colloidal suspensions have weaker shear-thinning with $\phi=0.36$ far from the sol-gel transition. The values of $\eta_{\infty}$ for all the volume fractions $\phi$ are given in Table-II. Fig.~\ref{fig:Rheometry}b represent dependence of loss modulus $G''$ and storage modulus $G'$ on shear strain $\gamma$ obtained from amplitude sweep experiments for various $\phi$ of colloidal and non-colloidal suspensions.

\section{Experiments}

\begin{figure*}[h!]
\includegraphics[width=\linewidth]{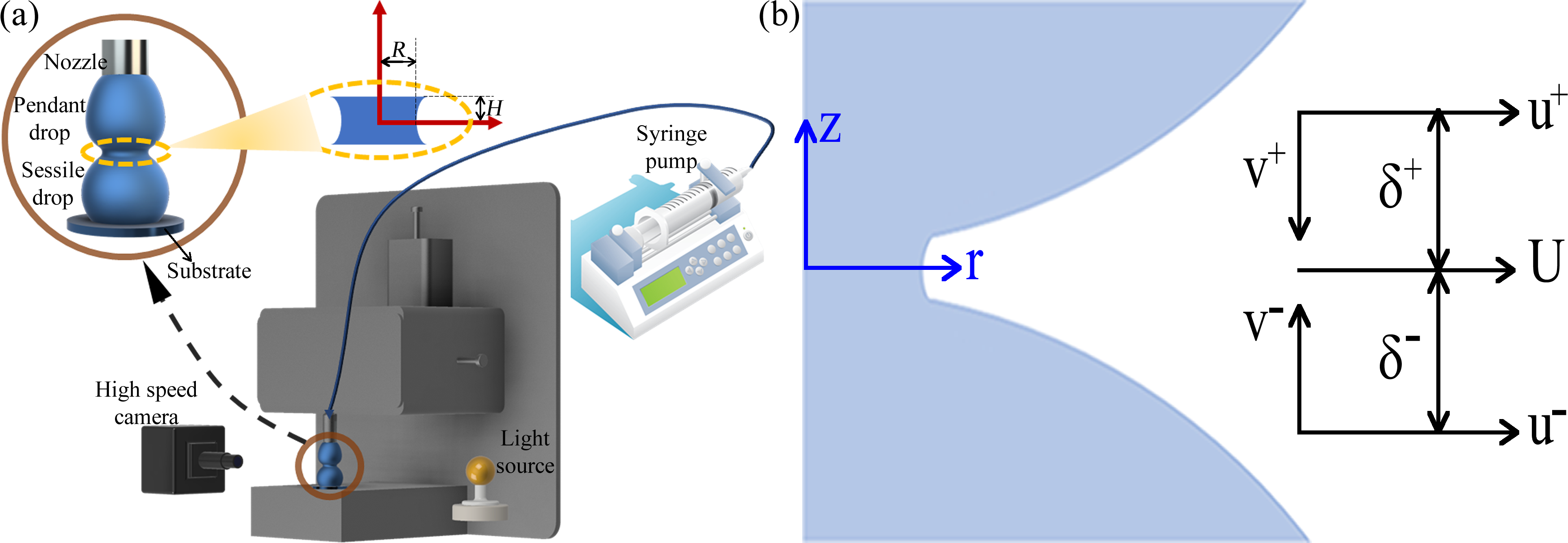}
\centering
\caption{(a) Schematic of experimental setup showing the geometric parameters neck radius $R$ and semi-bridge height $H$, (b) Schematic of flow at the neck region during initial stages of coalescence.}
\label{fig:SchematicSetup} 
\end{figure*}

A sessile drop is obtained by dispensing a droplet of volume $7.5 \mu l$ on a substrate. Coalescence is achieved by bringing a sessile drop of same volume towards the pendant drop at an approach velocity of  $10^{-4}$ m/s. The schematic of the experimental setup is shown in Fig.~\ref{fig:SchematicSetup}a. Bridge radius $R$ and semi-width $H$ are captured at 1,70,000 fps using high-speed camera (Photron Fastcam mini) along with a Navitar lens attachment. Temporal data on bridge shapes is tracked by a custom written edge detection algorithm with sub-pixel accuracy on MATLAB platform. Firstly, the background noise is removed using suitable thresholding intensity. The binary image so obtained gives the row coordinate of the contact point for the two droplets. Along this row, pixel intensities are investigated to find jumps in intensities which gives the location of relevant columns highlighting the bridge location. Further these coordinate points are used to obtain the neighbouring grey intensities from the original image. Based on these, sub-pixel coordinates for bridge location are obtained through five-point weighted averaging. These coordinates are then used to obtain the bridge radius $R$. Similarly at each time step, corresponding image is processed to obtain the temporal evolution data on bridge radius $R$.

Capillary breakup extensional rheometry dripping on a substrate (CABER-DOS) experimental setup and procedure is adapted from Dinic et. al.\cite{dinic2017pinch}. A pendant drop is created at the tip of a nozzle of radius $R_n=0.6$ mm by pumping at a flow rate of $Q=0.02$ mL/min. The pendant drop so obtained is deposited on a glass substrate placed below the nozzle tip at a distance of $H_d=6R_n$. Upon contact with substrate the droplet spreads creating a capillary between the nozzle tip and spreading fluid which is captured using high-speed camera. The temporal evolution data of capillary radius $R_c$ is obtained similarly as in coalescence with $t=0$ marked by the formation of the capillary bridge between the nozzle tip and spreading droplet.   
  
\section{Theory}

In order to develop a theoretical framework for power-law fluids, we employed the symmetry of sessile-pendant geometry for explicating the kinematics of flow field. The analysis is formulated in axis symmetric cylindrical coordinates under quasi-steady and quasi-radial assumptions as shown in Fig.~\ref{fig:SchematicSetup}b. Based on these the flow field $\textbf{V}=(V_r,0,V_z):=(u,0,v)$ can be simplified at the neck region $z=0$ as $u\ne 0$ and $v=0$. While for the small neighbourhood region of span $\delta$ on both sides, flow field can be simplified as $u^+=u^-$ and $v^+= -v^-$ respectively. As a consequence of these, terms in velocity gradient tensor at $z=0$ line can be simplified as $\frac{\partial u}{\partial z}=0$ and $\frac{\partial v}{\partial r}= 0$.   

For the coalescence geometry, conservation laws of mass and momentum can be simplified along the $z=0$ line as written in Eq(2) and Eq(3) respectively. The terms corresponding to stress tensor $\boldsymbol{\tau}$ are obtained by employing the Ostwald-de Waele power-law model constitutive behavior\cite{bird} represented in Eq(4).

\begin{equation}
\frac{\partial u}{\partial r}+\frac{u}{r}+\frac{\partial v}{\partial z}=0
\end{equation}
\begin{equation}
\rho\big( u\frac{\partial u}{\partial r}\big)=-\frac{\partial p}{\partial r} +\frac{\partial \tau_{rr}}{\partial r}+\frac{\tau_{rr}}{r}+\frac{\partial \tau_{rz}}{\partial z}\\
\end{equation}

\begin{equation}
\boldsymbol{\tau}=2\eta(\dot \gamma) \pmb{\dot D} \text{      where, } \pmb{\dot D} \text{ is the strain-rate tensor}
\end{equation}

By definition $\eta(\dot \gamma)=\kappa|\dot \gamma|^{n-1}$ with $\dot\gamma=\sqrt{\pmb{\dot D}:\pmb{\dot D}}=\sqrt{\Big(\frac{\partial u}{\partial r}\Big)^2 + \Big(\frac{\partial v}{\partial z}\Big)^2+\Big(\frac{u}{r}\Big)^2}$ being the second invariant of the strain-rate tensor, $\kappa$ the flow consistency index  and $n$ the degree of non-Newtonian behaviour with $n \in (0,1)$ representing shear thinning fluids and $n=1$, $n>1$ representing the Newtonian and shear thickening behaviour respectively. Owing to the quasi-steady and quasi-radial assumptions, $\boldsymbol{\tau}$ in Eq (4) can be reduced at $z=0$ line as $\boldsymbol{\tau}\sim 2^{\frac{n+1}{2}}\kappa\big|\frac{\partial u}{\partial r}\big|^{n-1}\pmb{\dot D}$ resulting in individual components of stress tensor $\boldsymbol{\tau}$ as:

\begin{subequations}
\begin{align}
\tau_{rr}\sim2^{\frac{n+1}{2}}\kappa\big(\frac{\partial u}{\partial r}\big)^n\\
\tau_{rz}\sim0
\end{align}
\end{subequations}

For obtaining the semi-analytical evolution equation of bridge radius $R$, the $r$-direction momentum equation Eq (3) is further simplified to Eq (6) through scaling arguments $u\sim U$, $r\sim R$, $z\sim \frac{R^2}{2R_o}$, $\frac{\partial p}{\partial r}\sim\sigma\big(\frac{1}{R^2}+\frac{2R_o}{R^3}\big)$ along with the components of stress tensor in Eq (3) as $\frac{\partial\tau_{rr}}{\partial r}\sim\frac{\tau_{rr}}{R}$ and $\frac{\partial\tau_{rz}}{\partial z}\sim\frac{\tau_{rz}}{\frac{R^2}{2R_o}}$ ($R_o$ is droplet radius and $\sigma$ is the surface tension). Here, Eq (6) is a polynomial equation with non-integral power of the form given in Eq (7) with $F(\kappa,\rho, R,n):=-\frac{2^{\frac{n+1}{2}}\kappa C_2}{\rho R^n}$ and $G(\sigma,\rho, R, R_o):=-C_1\frac{\sigma}{\rho}\Big(\frac{1}{R}+\frac{2R_o}{R^2}\Big)$ and $C_1$, $C_2$ being the scaling constants. The roots $K(\sigma,\rho,\kappa, R, R_o, C_1, C_2)$ of such equation are obtained by using the Newton-Raphson method with initial guesses coming from the special case of $n = 1$ where the roots are trivially obtained through quadratic formula. The final evolution equation for bridge radius $R$ is of the form given in Eq (8) where $U=\frac{dR}{dt}$.

\begin{equation}
\rho\Big(\frac{U^2}{R}\Big)=C_1\frac{\sigma}{R}\Big(\frac{1}{R}+\frac{2R_o}{R^2}\Big) + \frac{2^{\frac{n+1}{2}}\kappa C_2}{R}\Big(\frac{U}{R}\Big)^n
\end{equation}

\begin{equation}
U^2+F(\kappa,\rho,R,n)U^n+G(\sigma,\rho,R,R_o)=0
\end{equation}

\begin{equation}
\frac{dR}{dt}= K(\sigma,\rho,\kappa,R,R_o,n,C_1,C_2)
\end{equation}

First order finite difference scheme is used to obtain bridge radius $R$ from Eq (8). Time step $\Delta t=10^{-6}$ sec is taken sufficiently small to ensure numerical stability.

\section{Results and Discussion}

Coalescence in sessile-pendant configuration starts with the formation of liquid bridge of radius, $R$, and width, $H$, that is formed on proximate approach of the two droplets as depicted in Fig.~\ref{fig:SchematicSetup}a. The evolution of this bridge is driven by the interplay between capillary, viscous and inertial force to reach a thermodynamically stable state of single daughter droplet. Among these forces, capillary force assist the evolution dynamics while the other two oppose the process.     

\begin{figure*}[h!]
\includegraphics[width=\linewidth]{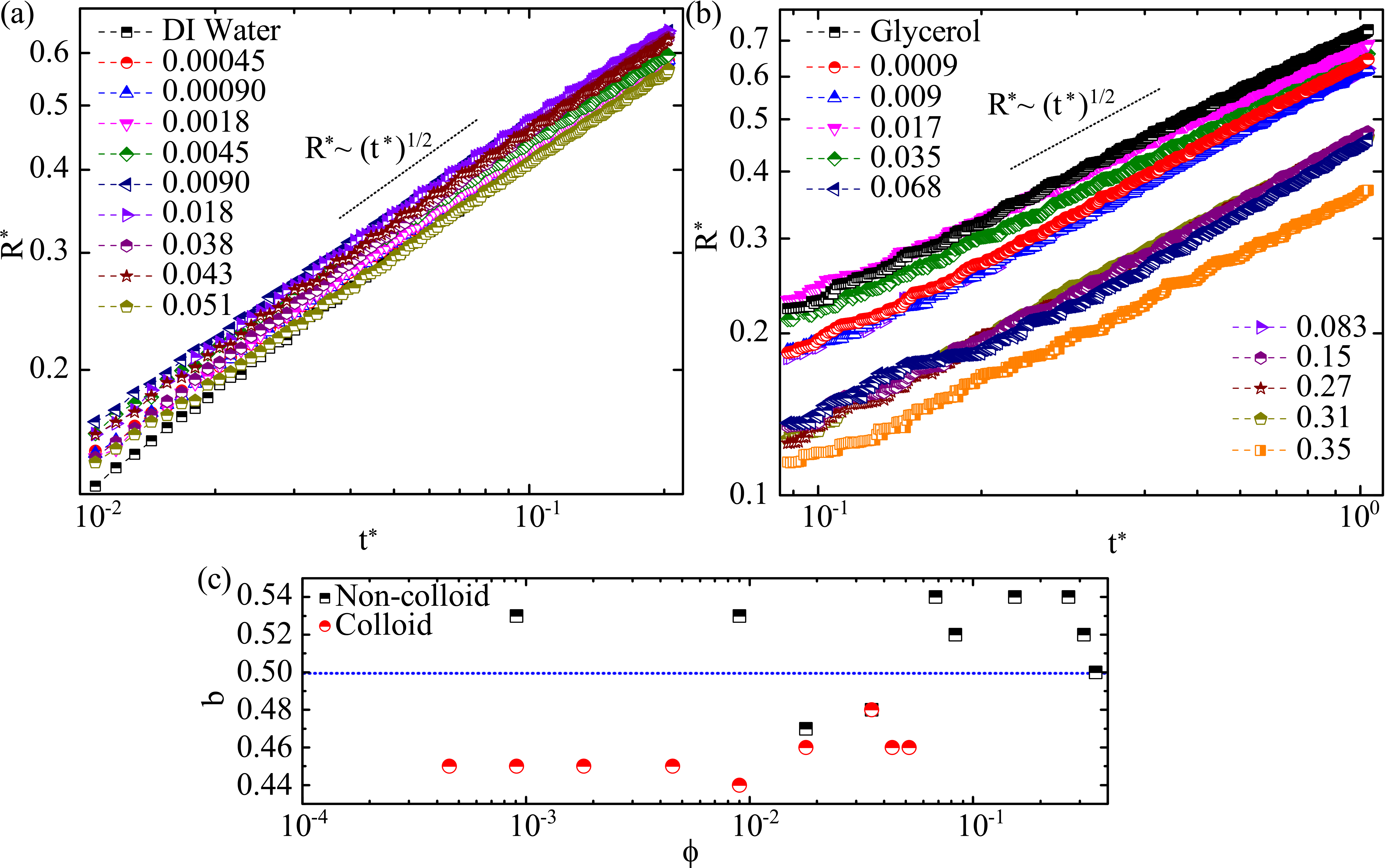}
\centering
\caption{Temporal evolution of neck radius showing a constant power law exponent $b$ of 1/2 for various $\phi$ of (a) Colloidal suspension, and (b) Non-colloid suspension  (c) Variation of exponent $b$ with $\phi$ in both colloidal and non-colloidal suspensions.}
\label{fig:Rvstdata} 
\end{figure*}

The temporal evolution of dimensionless liquid bridge radius $R^*=R/R_o$ with dimensionless time $t^*=t/t_i$ ($t_i=(\rho R_o^3/\sigma)^{0.5}$ is the inertial time scale), for various volume fractions of colloidal and non-colloidal suspensions, are represented in Fig.~\ref{fig:Rvstdata}a and Fig.~\ref{fig:Rvstdata}b respectively. The data shown for different volume fractions $\phi$ are an average of 4 trials. The error in the neck radius is $\pm5$\%. It is observed from Fig.~\ref{fig:Rvstdata}a and b that the bridge radius evolves as $R^*=a (t^*)^b$. For colloidal suspensions, it can be observed from Fig.~\ref{fig:Rvstdata}a that the evolution dynamics is independent of concentration $\phi$ and the power-law exponent, $b$, is equal to the universal Newtonian value $b=0.5$. However, for non-colloidal suspensions, the evolution is partially dependent on the concentration $\phi$. As observed from Fig.~\ref{fig:Rvstdata}b, the prefactor, $a$, decreases on increasing concentration but the $b$ remains $0.5$. For all the concentration $\phi$ represented in Fig.~\ref{fig:Rvstdata}a and b, the error in the measurement of $b$ is $\pm5\%$ and the value of the same are presented in Fig.~\ref{fig:Rvstdata}c.    

\begin{figure}[h!]
\includegraphics[width=\linewidth]{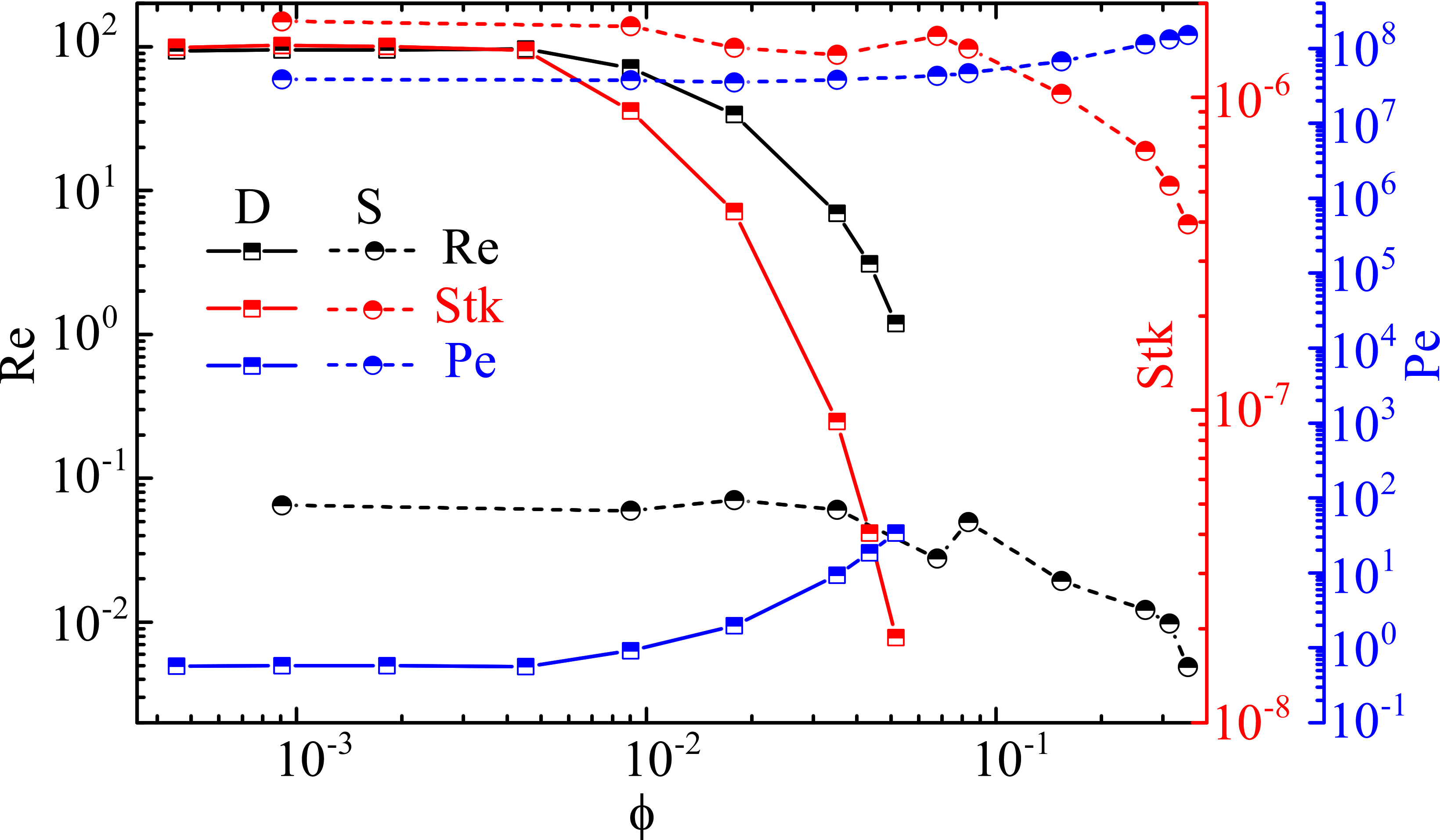}
\centering
\caption{Non-dimensional numbers, Reynolds number, Stokes number, and Peclet number calculated from experimental data at different $\phi$ for colloidal and non-colloidal suspensions (Note: D represents colloidal suspensions and S represents non-colloidal suspensions).
}
\label{fig:NonDimNumbers} 
\end{figure}

To expound the bridge evolution further, it is important to look at the relative magnitudes of the dominant governing forces in the continuum. For coalescence, these forces are capillary force, $F_c$, that drives the bridge growth, inertial force, $F_i$,  and viscous force, $F_v$, that oppose the evolution dynamics. For colloidal and non-colloidal suspensions, Brownian force, $F_b$, can also play a crucial role. The relative interplay of these forces is elucidated through Reynolds number, $Re=<\frac{\rho UL_c}{\eta_{\infty}}>$, Stokes number, $Stk=<\frac{\rho_p U d_p^2}{18\eta_{\infty} L_c}>$, and Peclet number, $Pe=<\frac{U d_p^3 \eta_{\infty}}{L_ckT}>$ where, $L_c\sim R$ and $U\sim \partial R/\partial t$ are the characteristic length and velocity scale associated with the flow respectively, $k$ is the Boltzmann constant, $\eta_{\infty}$ is infinite shear viscosity, and $T=300$ K is the absolute Temperature. The variation of  $Re$, $Stk$, and $Pe$ with concentration $\phi$ is represented in Fig.~\ref{fig:NonDimNumbers}. It is observed from the figure that the $Re$ for colloidal suspensions is of order greater than $\mathcal{O}(10^0)$ indicating that coalescence is in the inertial regime. While for non-colloidal suspensions, $Re$ is of $\mathcal{O}({10^{-2}})$ implying  viscous as the dominant force. However, the flow is in the inertial dominant region owing to the experimental time, which is in the range of inertial time scale $\tau_i=(\rho R_o^3/\sigma)^{0.5}\sim\mathcal{O}(1)$ ms. Further observation from Fig.~\ref{fig:NonDimNumbers} suggests that the Stokes number $Stk$ is of order less than $\mathcal{O}(10^{-6})$
for all the volume fractions $\phi$ for colloidal and non-colloidal suspensions suggesting that the particle relaxation time is much smaller than the characteristic time of the flow field. Therefore, for all the experiments particles were relaxed to the external flow field and were not creating extra resistance through slipping and local agglomeration in flow field which could potentially demonstrate shift from Newtonian coalescence. The relative magnitude of the Brownian forces represented by Pe is highlighted in  Fig.~\ref{fig:NonDimNumbers}. The Peclet number $Pe$ for colloidal suspensions is more than $\mathcal{O}(10^0)$ and for non-colloidal suspensions it is of $\mathcal{O}(10^6)$ signifying that inertial forces are always greater than the Brownian forces and therefore the effect of thermal forces can be neglected in the time scale of experiments.

\begin{figure*}[h!]
\includegraphics[width=\linewidth]{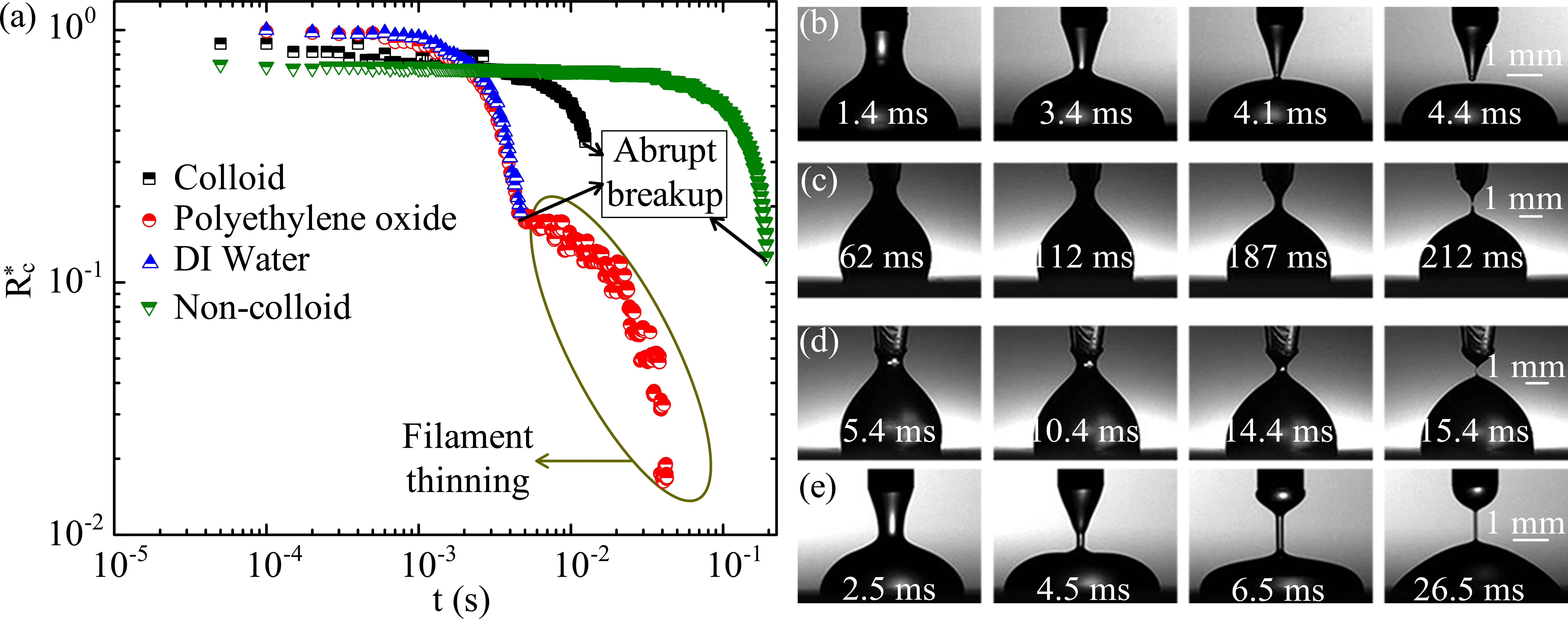}
\centering
\caption{(a) Temporal behaviour of the dimensionless capillary bridge radius of DI water,colloidal and non-colloidal suspensions, polyethylene oxide (PEO) representing the difference in the breakup of the filament. Snapshots showing the thinning dynamics at various instants of time for (b) DI Water, (c) colloidal suspensions, (d)non-colloidal suspensions, and (e) 0.1\% g/mol PEO of molecular weight $M_w=5\times10^6$ g/mol.}
\label{fig:CaberExpt} 
\end{figure*}

In order to develop a theoretical framework for understanding coalescence in colloidal and non-colloidal suspensions it is important to look at the rheological properties of these fluids. Fig.~\ref{fig:Rheometry}a and b represent the viscosity and amplitude sweep plots for the volume fractions $\phi$ used in the experiments. It can be observed from Fig.~\ref{fig:Rheometry}a that the colloidal suspensions are strongly shear thinning with their viscosity's changing significantly in orders as we move to higher volume fractions indicating an approach to a sol-gel transition. However, the non-colloidal suspensions are weakly shear thinning, and their viscosity doesn't change drastically with concentration indicating a higher concentration for sol-gel transition. Fig.~\ref{fig:Rheometry}b probes the visco-elasticity of respective fluids. As observed all volume fractions $\phi$ of colloidal suspensions have storage modulus $G'$ more than the loss modulus $G''$ implying they have significant elasticity to be classified as visco-elastic fluids. But non-colloidal suspensions have loss modulus $G''$ more than the storage modulus $G'$ implying that they are weakly viscoelastic and their behavior is much closer to Newtonian fluids. However, it is important to note that the viscosity and amplitude data are obtained through shear rheology therefore they may not be adequate to understand coalescence which is an extensional dominant flow. CABER-(DoS) experiments are performed to get insights into the extensional rheology of colloidal and non-colloidal suspensions. Fig.~\ref{fig:CaberExpt}a represents the temporal evolution of dimensionless capillary bridge radius $R_c^*=R_c/R_n$ (Here $R_c$ and $R_n$ are the capillary bridge and nozzle radius respectively) for DI Water, non-colloidal suspension $\phi=0.35$, colloidal suspension $\phi=0.043$ and Polyethylene oxide $M_w=5\times10^6$ $c=0.1\%$ g/ml. It is observed that the $R_c$ in colloidal and non-colloidal suspensions have similar evolution characteristics as DI water. The capillary abruptly breaks after initial universal thinning, implying negligible extensional elasticity. However, for polymer, the initial bridge thinning is followed by the formation of a filament that diminishes in time with characteristic relaxation time $\lambda$ implying finite extensional elasticity. Snapshots of such evolution of capillary bridge for different time instances is shown in Fig.~\ref{fig:CaberExpt}b. Utilizing the above observations, we can identically claim that all volume fractions $\phi$ of colloidal and non-colloidal suspensions explored in the current study have negligible elasticity in extensional flows similar to Newtonian fluids.            

\begin{figure}[h!]
\includegraphics[width=\linewidth]{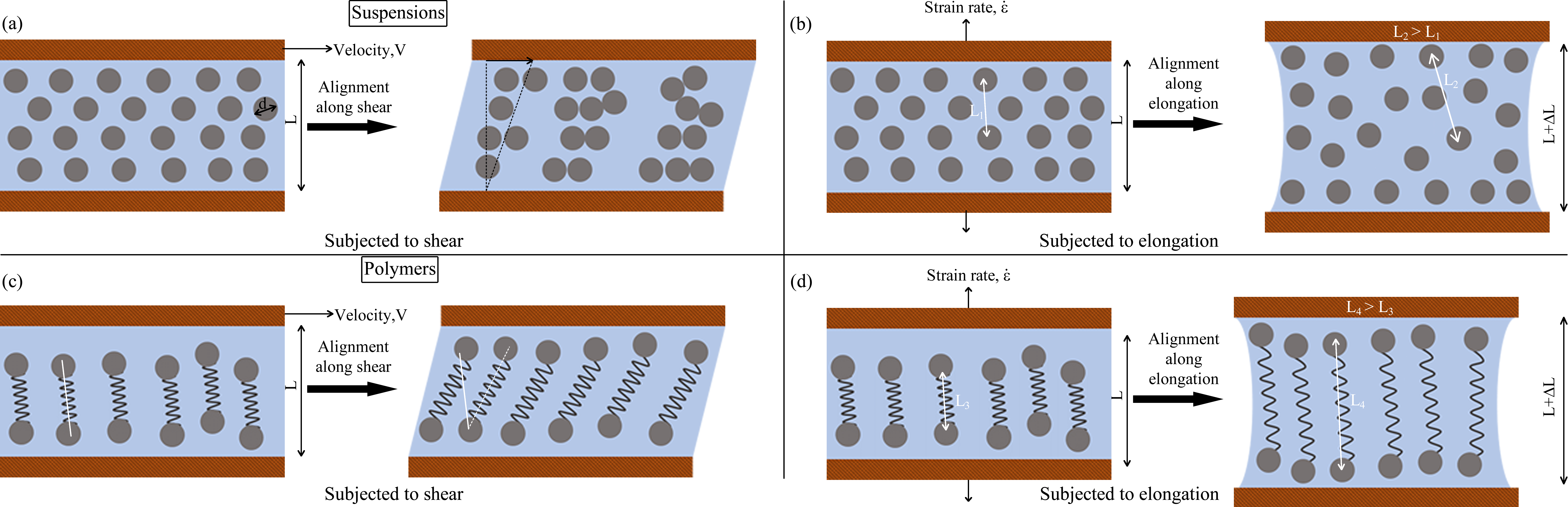}
\centering
\caption{Schematic representation of the behaviour of suspensions under (a) shear flow, (b) extensional flow. Flow behaviour of polymers subjecting to (c) shear flow, (d) extensional flow.}
\label{fig:SchematicMicrostructure} 
\end{figure}

Owing to this, the value of power-law exponent $b$ across volume fractions $\phi$ was found to be equal to the Newtonian exponent $b=0.5$ as illustrated by Fig.~\ref{fig:Rvstdata}c. Although these fluids have significant elasticity, unlike macromolecular fluids this elasticity is not reflected in the value of exponent $b$. This contrast in behaviour is the signature of difference in the micro-structures of the two subclass of complex fluids. As represented in Fig.~\ref{fig:SchematicMicrostructure}c and d, Macromolecuar fluids have large molecular chains that contribute to stresses in extensional and shear deformations. However, colloidal and non-colloidal suspensions as represented in Fig.~\ref{fig:SchematicMicrostructure}a have hard particles that only contribute to shear deformation through agglomerations. In extensional deformations Fig.~\ref{fig:SchematicMicrostructure}b, these particles are not able to interact as they are pulled apart from each other resulting in negligible extensional elasticity. This behaviour can be further understood from energy perspective. Coalescence in macromolecular fluids have a storage/dissipative dynamics with molecular chains absorbing part of the capillary energy while in suspensions it is dissipative alone with particles contributing to energy dissipation alone.

The rheological behaviour indicates that these dissipative fluids belong to the broader class of power law fluids and the theoretical framework developed above can be effectively used for understanding their coalescence. For the closure of the analytical solution, it is essential to study the physical behaviour of scaling parameters $C_1$,and $C_2$ represented in Eq (8). Among these scaling parameters, $C_1$ is the coefficient of capillary force that assists in bridge evolution while $C_2$ is the coefficient of axial stress $\tau_{xx}$ that resists the same. Owing to this nature, the coefficient $C_2$ will have a negative value while $C_1$ will have a positive value. The magnitudes of these scaling parameters are found by solving the equation for the special case of $n=1$ as this represents the special case of Newtonian fluids for which bridge radius $R$ is known to follow the universal dynamics of $R\sim t^{0.5}$ in the inertial dominant regime. The value obtained for $C_1$ and $C_2$ is $0.01$ and $-0.01$ respectively. It is important to note that our bridge evolution equation reduces to a similar form as was obtained by Xia et al.\cite{44} However, the two equations have different scaling constants owing to differences in characteristic length assumption.

\begin{figure*}[h!]
\includegraphics[width=0.82\linewidth]{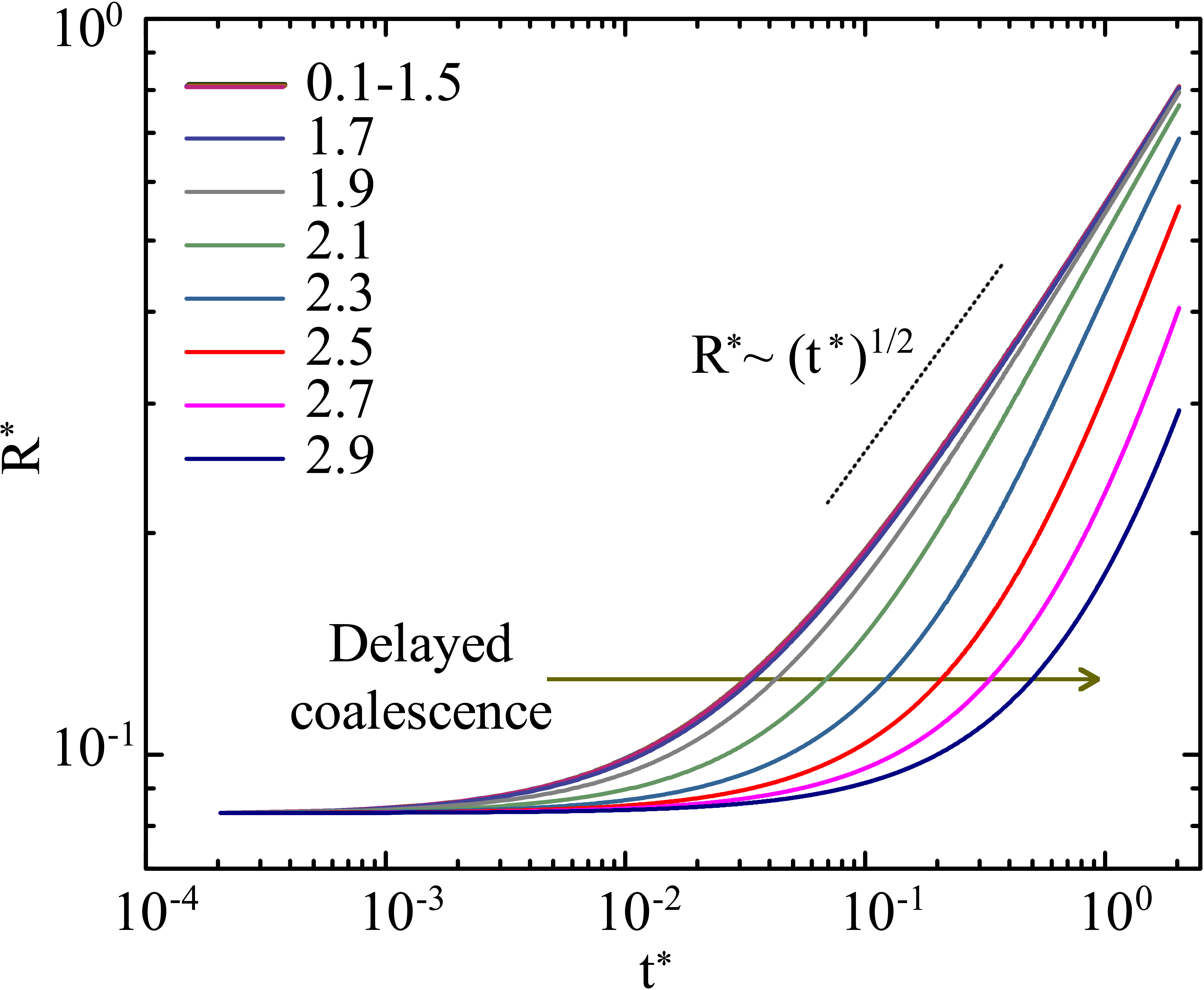}
\centering
\caption{Temporal evolution of neck radius obtained from solving Eq (8) for different power-law exponent $n$ in Ostwald-de Waele power-law model constitutive behavior.}
\label{fig:Theory} 
\end{figure*}

Physically the terms in Eq (6) corresponding to $U^2$ and $U^n$ are the dissipative terms representing inertia and viscous components respectively. The term corresponding to $\sigma$ represents the capillary component and is the driving force in coalescence dynamics. The dimensionless solution obtained on solving Eq (8) for different degrees of non-Newtonian behaviours $n$ at an interval of $0.1$ in $n\in(0,3]$ is represented in Fig.~\ref{fig:Theory}. It is observed that for thinning fluids $n<1$ the value of the power-law exponent is universal and equal to that of Newtonian fluids $n=1$. This is evident as in Eq (6), the term representing viscous contribution has degree $n<1$ making it lesser significant when compared with the inertial component having degree $2$. As a consequence inertial forces are the dominant governing force leading to the universal value of $b=0.5$. However for thickening fluids $(n>1)$ it is observed that the bridge radius $R^*$ starts to deviate from the Newtonian behaviour as $n$ increases. Such deviation in coalescence from Newtonian and thinning fluids is explained by the nature of Eq (6). As $n$ increases for thickening fluids, the viscous component starts to gain significance and at $n\geq 2$ it becomes the dominant term with the degree of Eq (6) being equal to $n$. However, it is important to note that this deviation is different from the sub-Newtonian deviation that occurs in macromolecular fluids. Unlike the latter, the coalescence in thickening fluid $n>1$ is delayed due to greater resistance offered by the viscous stresses instead of additional elastic stresses. Although Eq (6) captures the dynamics of thickening fluids qualitatively, experiments are required to validate the applicability of the model for them.

Further investigating Eq (6) we can look at a special case of arrested coalescence $b=0$. Eq (6) reduces to a quadratic polynomial of the form $U^2+FU+G=0$ at $n=1$. This polynomial always has real and at least one positive solution as its discriminant $F^2-4G>-F>0$ implying that for Newtonian fluids, coalescence is never arrested unlike those in macromolecular fluids. The following result can be simply extended to thickening and thinning fluids as the general function $H(U,F,G,p)=U^2+FU^p+G$ with $p\in \mathbb{R}$ will always have at least one real positive root as long as $H(U,F,G,1)=0$ satisfies the above conditions. The result is significant as it illustrates that arrested coalescence is the signature of elastic droplets. Therefore, if the microstructure of complex fluid droplets doesn't have a storage mechanism, they are always going to coalesce.

Overall our analysis highlights the importance of storage-based micro-structure for the existence of arrested coalescence via a sub-Newtonian transition regime. The experimental and theoretical observations have generalised the Newtonian universality of $b=0.5$ to complex fluids of subclass colloidal and non-colloidal suspensions that have thinning and zero extensional elasticity. The analysis has also explored thickening fluids where the coalescence appears to be delayed due to large resistive forces. 

\section{Conclusion}

In this work, we experimentally probe coalescence in colloidal and non-colloidal suspensions. Our observations have extended the Newtonian universality of power-law exponent $b=0.5$ to these complex fluid. The results are strengthened by the theoretical framework developed based on the Ostwald-de Waele constitutive law. The study highlights the importance of extensional elasticity for the existence of sub-Newtonian regime. colloidal and non-colloidal suspensions have micro-structures that are dissipative with non-zero shear elasticity but zero extensional elasticity resulting in Newtonian type coalescence. Further the theoretical framework is utilized to propose the non-existence of arrested coalescence in generalized Newtonian fluids with zero extensional elasticity. However, the current experiments and analysis neglect the effect of surrounding fluids and higher approach velocities between droplets by assuming air as the outer fluid and low approach velocity respectively. Further studies are required to understand the effect of surrounding fluids and higher approach velocities. The volume fractions explored in this study are thinning and far from sol-gel transition. Therefore future works are required to investigate highly concentrated colloidal and non-colloidal suspensions that have thickening and finite extensional elasticity.

\begin{acknowledgement}
A.K. acknowledges partial support from DST-SERB, A.S.R., and S.C.V. acknowledge partial support from the MHRD.
\end{acknowledgement}

\bibliography{Manuscript}
\end{document}